\documentclass[aps,prl,showpacs]{revtex4-1}
\usepackage{amssymb}
\usepackage{amsmath}
\usepackage{graphicx}
\usepackage{epsfig}
\begin{document}

\title{Nonlinear non-signaling Schr\"odinger equation}
\author{Tam\'as Geszti}
\affiliation{ Department of the Physics of Complex Systems, \\ E\"otv\"os University; H-1117 Budapest, Hungary 
\\e-mail: tamas.geszti@ttk.elte.hu}
\date{01 March 2024}
\begin{abstract} 

In looking for new insight into the nature of the quantum-classical border still unclear after a century of research, a new kind of nonlinear extension of Schr\"odinger's wave equation is proposed. It ensures non-signaling by keeping linear the evolution of {\it{coordinate-diagonal}} elements of the density matrix. The nonlinear part of the equation includes a negative kinetic energy operator characterized by a universal critical mass $\mu$, estimated to be about  $2\cdot10^{-23}~$kg. As some effective mass $M$ belonging to a pointer variable $x$ grows beyond $\mu$, spreading of wave packets along $x$ turns into its reverse: collapsing, which marks the quantum-classical border. Interference of large molecules is suggested for an experimental check of the proposed framework. Phase differences among terms of a prepared superposition are forced by the non-signaling condition to be preserved during collapse; a possible reason for the observed remote correlations.
\end{abstract}


\maketitle

The nature of the quantum-classical border is one of the great unsolved questions of 20-21th century physics, with various attempts competing, without any broadly accepted outcome so far. Atoms, molecules and any kind of groups of subatomic particles entangled by coming from a common source or just once interacted, move like a single  wave in their configuration space, accurately described by Schrödinger's linear wave equation. That explains many properties, including spectra, chemical bonding, various kinds of interferometry to an accuracy limited by our measuring and calculating abilities only: wave-like motion is a fact. However, to observe those waves, one needs some very special tool: detectors, which are metastable systems, eventually flipped over by arriving waves of atoms, electrons, photons etc. to give a signal of visible size. Then, as demonstrated by a huge amount of experiments, nonlinear phenomena emerge: as soon as a superposition would grow into macroscopically different terms - the famous "Schrödinger's cat" - it collapses into an apparently randomly chosen term of it. More than that, during measurement - in a so far unknown way - the incoming entanglement is translated into strange correlations between remote events, often reminding of some compact object - particle - going one way or another. Although everyday philosophy might suggest that those correlations should be just consequences of the common source acting as common cause, the necessarily nonlinear dynamics describing the emergence of remote correlations could not be found during many decades, giving ample space to fantasies dominated by the Einstein bonmot "spooky action-at-a-distance". This Letter offers a way to see how the common cause scenario can work, tracing back remote correlations to entangled initial conditions, followed by fully local nonlinear dynamics \cite{Griffiths}.

History of the topic has long been dominated by an extremely successful phenomenology: the Copenhagen interpretation, connecting Schrödinger's wave function to measurement statistics through the shortcut of Born's rule, at the price of admitting to regard the wave function just a tool to calculate, not a real physical quantity. That describes the observed statistics with high accuracy, down to refined properties like all kinds of experiments violating Bell-CHSH and GHZ inequalities \cite{Brunneretal,ChshGhz}. However, it opens up the way towards  mysteries like parallel worlds, or the idea of collapse being a consequence of looking at the detector, or the cat; besides, attributing physical reality to the seeming particles is a source of apparent backwards-in-time causal influences \cite{Aharonov}. From a more philosophical viewpoint, restricting ourselves to statistics, and getting forced to give up any insight to the enormous variety of individual events happening through wavelike motion, is a legitimate source of dissatisfaction \cite{EprPbr}. Admitting nonlinearity would offer hope to find an escape from all that, clarifying the dynamical origin of quantum randomness, and gaining insight into the whole dynamical background of the quantum-classical transition, including Born's rule; last not least, offering an explanation to the mysterious-looking remote correlations. 

Beside a great variety of stochastic models that - by postulating some kind of inherent randomness - offer a lot of insight beyond the Copenhagen interpretation \cite{collapse}, there have been various attempts to find valid nonlinear extensions to Schrödinger-type wave equations \cite{Bialynicki,Weinberg,Polchinski,Elze,Vecsi}, some exploring hydrodynamic analogies centered around a kind of continuum equation \cite{Doebner,Richardson,Sbitnev} that might offer some taste of locality. Although the proposed equations remain at the nonrelativistic limit, a strong relativity-based requirement appeared soon that has proved extremely useful in restricting the possibilities: one should avoid the possibility of "signaling", i.e. influencing remote events by locally imposed changes acting at a distance in no time; in the language of relativity: at superluminal speed \cite{Gisin}. General as that requirement sounds, it appeared immediately in a technically useful form: noticing that signaling scenarios typically build upon manipulating linear combinations of nonlinearly evolving density matrices, one concludes that nonlinear dynamics of the wave function can still be admitted, if a much looser necessary condition is obeyed: one should preserve linear evolution of the density matrix
\begin{equation}\label{dnstmtrx}
 \varrho\left(x, x^{\prime}, t\right)=
\Psi(x, t)\Psi^{*}(x^{\prime},t),  
\end{equation}
where $\Psi(x, t)$ is a pure-state wave function, depending on time 
$t$ and configuration space vector $x$, the multi-local components of which include the full microscopic coordinates of all detectors involved in the measurement process, along with all their environments  \cite{Tsirelson}. Although that approach looks a highly promising way to follow, to take that escape from the catch of signaling by any deterministic nonlinear wave equation has been demonstrated to be impossible \cite{BassiHejazi}. The present note is about opening a way round that final obstacle, by taking one more step back, and admitting a relieved version of the requirement of density matrix linearity that makes the task solvable.

Our starting point is the observation that whenever using the density matrix (\ref{dnstmtrx}) to calculate the mean value of any local operator in the configuration space, toward the end one always takes the limit $x^{\prime}\to x$. Accordingly, to grant non-signaling one need not require linear evolution of the whole density matrix; it suffices to assure linear evolution of its {\it coordinate-diagonal} elements 
$\lim_{x^{\prime}\to x}\varrho\left(x, x^{\prime}, t\right)$.

To turn that into a practical requirement, we go back to definition (\ref{dnstmtrx}) and introduce the upper or lower index "${n l}$" as a shorthand notation for nonlinear contributions to a time derivative. Then the limit $x^{\prime}\to x$ can be taken directly, presenting the  non-signaling condition in the final form
\begin{equation}\label{nonsig}
\lim_{x^{\prime}\to x}
\partial_{t}^{n l}\left(\Psi(x, t) \Psi^{*}(x^{\prime}, t)\right)=
\Psi^{*} \dot{\Psi}_{n l}+\Psi\dot{\Psi}_{n l}^{*}= 0,
\end{equation}
 a just-strong-enough necessary condition, which is mild enough to be compatible with a deterministic nonlinear wave equation. 

Looking for an appropriate nonlinear extension of Schrödinger's equation that satisfies the above condition, we observe that $\Psi^{*} \dot{\Psi}_{n l}$ and $\Psi\dot{\Psi}_{n l}^{*}$ are complex conjugate quantities adding up to zero, therefore - in accordance with Eq. (\ref{nonsig}) - they should satisfy a pair of conjugate equations
\begin{equation}\label{conjugate}
 \Psi^{*} \dot{\Psi}_{n l}=-i~ \mathbb{K}(\Psi,\Psi^{*}); \qquad
 \Psi\dot{\Psi}_{n l}^{*}=i~ \mathbb{K}(\Psi,\Psi^{*}), 
\end{equation}
containing the same real-valued quantity $\mathbb{K}(\Psi,\Psi^{*})$ to 
be specified by further requirements. Obviously, it should be nonlinear to admit collapse; besides, to avoid any by-pass appearance of signaling, it should be kept fully local, excluding any kind of integral or any scalar product of operators. 

To step further, some preliminary idea is needed about the way collapse may happen. As a so far unexplored territory, we choose this: collapse is the reverse of the well-known spreading of a wave packet, which is driven by the kinetic energy, and can be reversed by adding some kind of negative kinetic energy $(+\hbar^2/2\mu)\Delta$ with a characteristic mass $\mu$ \cite{negkin}.  A simple real-valued, local expression, nonlinear in $\Psi(x, t)$  and $\Psi^{*}(x, t)$, including that negative kinetic energy is this:
\begin{equation}\label{Kdef}
    \mathbb{K} := \frac{\hbar}{4 \mu}
    \left(\Psi \Delta \Psi^{*}+\Psi^{*} \Delta \Psi\right),
\end{equation}
where the strength of nonlinearity is characterized by the constant $\mu$, postulated as a universal mass marking the quantum-classical border. Linear quantum mechanics is valid for masses much smaller than $\mu$; nonlinearity is taking the lead as an effective mass $M$ appearing in the linear kinetic energy $(-\hbar^2/2 M)\Delta$ is growing too large.

Accepting that choice, and including a linear Hamiltonian $H_{lin}$ carrying all relevant properties of the system actually studied, we arrive at the conjugate pair of nonlinear Schrödinger equations 
\begin{equation}\label{Schrnl}
\dot{\Psi}=-\frac{i}{\hbar}\left(H_{lin} \Psi+\frac{\hbar^{2}}{4 \mu}\left(\frac{\Psi}{\Psi^{*}} \Delta \Psi^{*}+\Delta \Psi\right)\right);  \quad 
\dot\Psi^{*}=\frac{i}{\hbar}\left(H_{lin} \Psi^{*}+\frac{\hbar^{2}}{4 \mu}\left(\frac{\Psi^{*}}{\Psi} \Delta \Psi+\Delta \Psi^{*} \right)\right)
\end{equation}
interconnecting the dynamics of the conjugate waves $\Psi(x, t)$  and $\Psi^{*}(x, t)$, to be used in the rest of the present study \cite{normpreserving}. 

As a first technical check, we mention that for remote, uncorrelated objects labeled by some index $\alpha$, the wave function is in a direct-product state 
\begin{equation}\label{separated}
\Psi(x, t)=\prod_{\alpha}\psi_{\alpha}\left(x_{\alpha}, t\right).
\end{equation}
Since in that case the ratio $\Psi / \Psi^{*}$ is a direct product too, and
\begin{equation}\label{deltaseparate}
\Delta=\sum_{\alpha} \Delta_{\alpha}
\end{equation}
is a sum over the configuration space dimensions belonging to the different remote objects, Equations (\ref{Schrnl}) are separated into local ones - a basic requirement towards nonlinear extensions of Schrödinger's equation.

The above equations apparently entail a continuity equation \cite{Doebner,Richardson,Sbitnev} as required by matter conservation \cite{normpreserving}. There is an important bonus to that: although non-signaling would allow some nonlinear current of zero divergence, here the conserved current is fully due to linear quantum dynamics, with no contribution from nonlinearity, since obviously $(\Psi^{*}\nabla\Psi-\Psi\nabla\Psi^{*})+(\Psi\nabla\Psi^{*}-\Psi^{*}\nabla\Psi)$ is vanishing, not only its divergence. Luckily, this excludes violation of mass additivity across the quantum-classical border, empirically known to be valid from atomic to macroscopic scales, ever since the 19th-century uprise of atom-based chemistry, linked to the discoveries of Dalton and Avogadro. To violate that rule would be another possible harmful side effect of nonlinear quantum dynamics; our approach avoids that catch too. 

As mentioned above, the nonlinear terms in  Eqs. (\ref{Schrnl}) contain the Laplacian divided by a positive $\mu$ acting as a kind of negative kinetic energy, piloting wave packets towards collapse. As a basic example, take a particle detector with an avalanche of growing effective mass $M$ that moves along some "pointer variable" $x$ under driving force $-\nabla V(x)$; then a $1 \mathrm{D}$ toy model is described by an effective wave function $\Psi(x,t)$, obeying the nonlinear Schrödinger equation
\begin{equation}\label{toy}
 \dot{\Psi}=-\frac{i}{ \hbar}\left( -\frac{\hbar^2}{2M} \Delta \Psi
 + V(x)\Psi +\frac{\hbar^2}{4 \mu}\left(\frac{\Psi}{\Psi^{*}} \Delta \Psi^{*}+\Delta \Psi\right)\right).   
\end{equation}
As $M$ outgrows $\mu$, nothing happens to the nonlinear part of the Hamiltonian; it is the {\sl linear} kinetic energy $-\left(\hbar^{2} / 2 M\right) \Delta$ which is going negligible. As a consequence, the well-known spreading of wave packets, driven by quantum kinetic energy under the control of Coulomb forces over the whole microscopic quantum world, gradually disappears; instead, its reverse: shrinking takes the lead. 

Nonlinearity enters Eqs. (\ref{Schrnl}) in the form of the ratio $\Psi/\Psi^{*}$. The Madelung-Bohm representation \cite{Mad}
\begin{equation}\label{mad}
\Psi=A \exp (i \varphi)    
\end{equation} 
arranges conjugate pairs $\left(\Psi,\Psi^{*}\right)$ according to  "hyperbolic coordinates" $A=\sqrt{\Psi\Psi^{*}}$ and $\varphi=-i\ln{\sqrt{\Psi/\Psi^{*}}}$. The present approach is focusing on the active dynamical role of phase, excluding any direct change of the amplitude by nonlinearity, thereby satisfying the non-signaling requirement. In this representation Eq. (\ref{toy}) is displayed in two coupled equations: a continuity equation
\begin{equation}\label{contin}
\partial_{t} A^{2}=-\vec{\nabla} \cdot \mathbf{j};  \qquad  \mathbf{j}~=~ A^{2}~ \frac{\hbar \vec{\nabla} \varphi}{M}, 
\end{equation}
which - as required by non-signaling - contains no nonlinear contribution to the current, and an equation for the evolution of the phase $\varphi(x, t)$,
\begin{equation}\label{phase}
\dot\varphi=\frac{\hbar}{2}\left(-\frac{1}{M}+\frac{1}{\mu}\right)\left(-\frac{\Delta A}{A}+|\nabla \varphi|^{2}\right)-\frac{1}{\hbar} V.
\end{equation}
A striking thing to notice in the above couple of equations is that although the conserved current $\mathbf{j}$ as it appears in Eq. (\ref{contin}) contains no nonlinear part, its time derivative does have a nonlinear contribution through phase dynamics, Eq. (\ref{phase}). That is a hint that the measurement process is an interplay between nonlinear and linear dynamics, and visible collapse happens in two steps: first changing phases with active contribution from nonlinearity, then - as a consequence - amplitudes, changed by the linear currents induced by phase changes. That seems to follow the Lagrangian two-step way to Newtonian dynamics: coordinates determine forces that change momenta, finally influencing coordinates. 

In Eq. (\ref{phase}) the quantum-classical boundary appears as a kind of phase transition, taking place as soon as its control variable: the growing effective mass $M$ is crossing the critical value $\mu$. In particular, the r.h.s. of Equation (\ref{phase}) suggests an inspiring analogy between quantum collapse and the roughening of growing surfaces \cite{Rácz,Vicsek}, both being driven by a term like $|\nabla \varphi|^{2}$ as soon as its coefficient turns positive, making smoothness unstable in a noisy environment \cite{KPZ}. As for collapse, that noisy environment may turn out to be a back-action of the "einselection" process \cite{Zurek,Paternostro}.

Now we turn to check the chances of finding evidence for the nonlinear scenario introduced here in non-measurement situations. This starts by an order-of-magnitude estimate of the critical mass $\mu$; for that - following Ref. \cite{Geszti2018} - we consider the mass of the growing electron avalanche desired to produce a clear signal by an avalanche photon detector. According to review \cite{photondetector}, that requires an avalanche of $2\cdot10^7$ electrons, the mass of which is $\mu = 2\cdot10^{-23}~$kg; that is our first estimate of the critical mass appearing in the above formalism to mark the quantum-classical border. It is 15 orders of magnitude smaller than the Planck mass, which seems to exclude any direct hint to gravitational origin \cite{grav}; it is just a measure of nonlinearity, marking the borderline of the microscopic quantum world. Looked from the other side: the present framework has a good chance to build up peaceful coexistence with general relativity, like any other approach in which randomness is emergent from deterministic-chaotic dynamics.

Experiments approach the borderline drawn by the above estimate of mass $\mu$ from both sides: nanomechanical resonators from above, molecular interference from below. As for nanomechanics \cite{NANO}, doubly clamped carbon nanotube resonators now reach the lowest mass of $10^{-21}$ kg, not too promising to look for quantum nonlinearity, especially when you consider the side-effects of mechanical nonlinearities. The lower side looks more interesting: molecules of $10^{-25}$ kg have been demonstrated to interfere \cite{Arndt}, mentioning that to detect interference of molecules heavier by a factor of 100 looks technically feasible, however, so far not yet confirmed. That seems to be a hot territory to look for effects of quantum nonlinearity of the kind suggested here; the actual effect may be an uprising tendency to turn spreading into shrinking, that may block the beam widening necessary to observe few-slit interference phenomena.

Although the above suggestion to a large-molecule interference test is the main result of the present note, the way the non-signaling condition (\ref{nonsig}) influences the dynamics described by Eqs. (\ref{Schrnl}) is worth more discussion. The main philosophy is this: while incoming entanglement is converted by nonlinear dynamics into outcome statistics, non-signaling acts as a bunch of hard restrictions, forbidding to forget incoming data. That is an extended variant of Pearle's  Martingale principle \cite{Pearle}, serving as a way to explain remote correlations through preserved input, with no action-at-a-distance.
For a first insight into the way all that can be hosted by our mathematical framework, consider an arrangement of incoming waves from the same source, undergoing some kind of Stern-Gerlach-type separation (also called "Von Neumann measurement"). Subsequently each of the respectively placed detectors meets its partial beam, thereby the detectors become part of the entangled state
\begin{equation}\label{entgld}
   \Psi=\sum_k u_k =\sum_k \prod_\alpha u_{k,\alpha} , 
\end{equation}
where $u_{k,\alpha}$ is the local state of the $\alpha$'th detector locally entangled with the arriving partial beam, in the $k$'th term of the fully entangled state. Substituting that structure as an Ansatz into the sum and difference of Eqs. (\ref{conjugate}) and using the definition (\ref{Kdef}), one obtains two equations, one for the non-signaling condition, the other for the dynamics permitted by the first:
\begin{equation}
    \begin{split}
        \sum_k\sum_l\sum_{\alpha}\prod_{\beta\not=\alpha}&u^{*}_{k\beta}u_{l\beta}
        \bigl(u^{*}_{k\alpha}\partial_t^{nl}u_{l\alpha}
        +u_{l\alpha}\partial_t^{nl}u^{*}_{k\alpha}\bigr)~=~0;
        \\
        \sum_k\sum_l\sum_{\alpha}\prod_{\beta\not=\alpha}u^{*}_{k\beta}u_{l\beta}
        \bigl(u^{*}_{k\alpha}\partial_t^{nl}&u_{l\alpha}
        -u_{l\alpha}\partial_t^{nl}u^{*}_{k\alpha}
        +i\frac{\hbar}{2\mu}(u^{*}_{k\alpha}\Delta u_{l\alpha}
        +u_{l\alpha}\Delta u^{*}_{k\alpha})\bigr)            ~=~0.
    \end{split}
\end{equation}
The above equations are apparently nonlocal in view of the nonlocal weights $u^{*}_{k\beta}u_{l\beta}$ with $\beta\not=\alpha$. However, the structure offers an escape from the catch: if at any place $\alpha$, any pair of local components $(u^{*}_{k\alpha},u_{l\alpha})$ satisfies a pair of local equations:
\begin{equation}\label{localnonsig}
 \partial_t^{nl} \ln(u^{*}_{k\alpha} u_{l\alpha})~=~0
\end{equation}
for the non-signaling condition, and
\begin{equation}\label{localdyn} 
 \partial_t^{nl} \ln\Bigl(\frac{u^{*}_{k\alpha}}{u_{l\alpha}}\Bigr)=                 i\frac{\hbar}{2\mu}\Bigl(\frac{\Delta u^{*}_{k\alpha}}{u^{*}_{k\alpha}}
 + \frac{\Delta u_{l\alpha}}{u_{l\alpha}}   \Bigr)
\end{equation}
for the dynamics, then what happens at location $\alpha$ is going on independently of whatever happens at different locations: "spooky action-at-a-distance" has been excluded. Eq. (\ref{localnonsig}) immediately carries one more important point: by the local Madelung-Born representation $u_{k\alpha} = a_{k\alpha} \exp_{k\alpha} (i \varphi_{k\alpha})$ it is turned into two equations:
\begin{equation}
   \partial_t^{nl} \ln(a_{k\alpha} a_{l\alpha}) ~=~0; \qquad
   \dot\varphi_k^{nl} - \dot\varphi_l^{nl}~=~0,
\end{equation}
the second of which is a particularly strong clue for the origin of remote correlations: non-signaling forces nonlinear evolution to preserve initially prepared phase differences among terms of the incoming superposition, much in the Martingale way \cite{Pearle}.

 Further details of the two-step dynamics appearing in Eqs. (\ref{contin},\ref{phase}) will be the subject of forthcoming research. In the meantime we mention that besides the routinely measurement- and non-measurement-based tests of various approaches to the problem of the quantum-classical border, any proposed solution may leave its footprint on several other phenomena.
An example that seems worth the closer investigation is local quantum scattering events leaving traces spreading in their environment towards the macroscopic scale, as happens e.g. in electric conduction; the slow dynamics of that can be a source of $1 / \mathrm{f}$ noise (flicker noise), a phenomenon explained in some special cases but not understood in its observed generality.

A less studied aspect of noise dynamics is this: detector bias thresholds are usually set empirically to eliminate noise that would violate Born's rule. Nonlinear quantum mechanics may have a chance to get insight into details of that.

Finally, the answer to "how does a detector know what the other is doing", is this: she does not know; they do correlated actions because they receive the same command from a common source, through the channel of entanglement. To analyze in details the way that happens, using the nonlinear extension of Schrödinger's equation proposed here, remains for forthcoming research. In the meantime, a useful analogy is this: spreading the same belief towards remote places may cause correlated actions of remote people who never had any personal contact among themselves. 

\section{Acknowledgments}
I am deeply indebted to Lajos Diósi for unlimited discussions about quantum foundations over many years. Many thanks to him and to Tamás Vicsek for carefully reading a preliminary version of this paper, with important suggestions to change. The present research had no financial support from any source.


\begin{thebibliography}{40}
\bibitem{normpreserving} This is norm-preserving evolution; unitarity, preserving all scalar products, is excluded by nonlinearity.
\end{thebibliography}
\end{document}